\documentclass[aps,prl,twocolumn,showpacs,preprintnumbers]{revtex4}
\usepackage{epsfig,epstopdf}
\usepackage{bm,graphicx,amsmath,amssymb}
\draft
\preprint{Submitted to Phys. Rev. Lett.}
\begin{document}
\newcommand{\bfi}{\it}
\newcommand{{\bfr}}{\mathbf{r}}
\newcommand{\beq}{\begin{equation} } 
\newcommand{\eeq}{\eeq{equation} } 
\newcommand{\comment}[1]{\vspace{1mm}\par
\framebox{\begin{minipage}[c]{.95 \textwidth} \rm #1
\end{minipage}}\vspace{1 mm}\par}
\newcommand{\rem}[1]{}
\newcommand{\remfigure}[1]{#1}
\newcommand{\neu}[2]{ \marginpar{\fbox{\color{red}\bf #1}}
                      {\color{blue}\bf #2} }
\newcommand{\rhobar}{\overline{\rho}}
\newtheorem{thm}{Theorem}
\preprint{}

\title{
Aggregation of finite size particles 
with variable mobility}
\author{ Darryl D. Holm${}^{1,2}$ and Vakhtang Putkaradze${}^{3}$}
\address{${}^{1}$Theoretical Division, 
Los Alamos National Laboratory, Los Alamos, NM 87545, USA
\\{\footnotesize email: dholm@lanl.gov} 
\\${}^{2}$Mathematics Department,
Imperial College London, SW7 2AZ, UK
\\{\footnotesize email: d.holm@imperial.ac.uk} 
\\${}^{3}$Department of Mathematics and Statistics, 
University of New Mexico,  Albuquerque, NM 87131-1141,
\\{\footnotesize email: putkarad@math.unm.edu}
}

\begin{abstract} \noindent
New model equations are derived for dynamics of self-aggregation of
finite-size particles. Differences from standard Debye-H\"uckel \cite{DeHu1923} and
Keller-Segel  \cite{KeSe1970} models are: the mobility of particles depends on the
configuration of their neighbors and linear diffusion acts on
locally-averaged particle density. The evolution of collapsed states
in these models reduces exactly to finite-dimensional dynamics of
interacting particle clumps. Simulations show these collapsed (clumped)
states emerge from smooth initial conditions, even in one spatial
dimension.

\noindent
{\bf Keywords:} gradient flows, blow-up,
chemotaxis, parabolic-elliptic system, singular solutions
\end{abstract}
\maketitle




%

Modeling finite size effects in the aggregation of interacting particles requires  
modifications of the class of Debye-H\"uckel equations. This problem is
motivated by recent experiments using self-assembly of nano-particles in
the construction of nano-scale devices \cite{XB2004}. Fundamental
principles underlying the self-assembly at nano-scales are non-local
particle interaction and nonlinear motion due to variations of
mobility at these scales.  
\rem{We shall keep our framework general
enough that our method of modeling nonlocal interactions among
different particles may be extended to the other physical
problems of interest described in the previous section. }
The model should account for the change of mobility due to the
finite size of particles and the nonlocal interaction among the
particles.  The local density (concentration) of particles is
denoted by $\rho$. For particles interacting pairwise via the 
potential $-G(|\mathbf{r}|)$, the total potential at a point ${\bfr}$ is 
$
\Phi({\bfr})=-\int \rho({\bfr} ') G(| {\bfr} - {\bfr}') \mbox{d} {\bfr}'=G * \rho 
\label{potential} 
$ 
where $*$ denotes convolution and $G>0$ for attracting particles. The
velocity of the particle is assumed to be proportional to the gradient of
the potential times the mobility of a particle, $\mu$.  The mobility can be
computed  explicitly for a single particle moving in an infinite fluid.
However, when several particles are present, especially in highly dense
states, the mobility of a particle may be hampered by interactions with
its neighbors. These considerations are confirmed, for example, by the
observation that the viscosity  of a dense suspension of hard spheres in
water diverges, when the density of spheres tends to its maximum
value.  Many authors have tried to incorporate the dependence
of mobility on local density by putting $\mu=\mu(\rho)$ and assuming
$\mu(\rho)\to0$ as $\rho\to\rho_{\hbox{\small max}}=1$
\cite{Velazquez2002}. Vanishing mobility
leads to the appearance of weak solutions in the equations, 
to singularities and, in general, to massive complications and
difficulties in both theoretical analysis and computational
simulations of the equations. 
\rem{In contrast, the Keller-Segel model specifies $\mu(\Phi)$, so the
mobility is taken in that case as a function of the concentration of
chemotactic agent (potential) $\Phi$.}

Alternatively, we suggest that the mobility $\mu$ should depend on an 
averaged density $\rhobar$ over some sampling volume, 
rather than on either the potential, or the exact
value of the density at a point. This assumption makes sense from the
viewpoints of both physics and mathematics. From the physical point of
view, the mobility of a finite-size particle must depend on the
configuration of particles in its vicinity. While attempts have
been made to approximate this dependence by using derivatives of
the local density, this approach may lead to unphysical 
negative diffusivity \cite{Villain1991}. 
Hence, we assume instead that the local mobility
depends on an integral quantity $\rhobar$ which is computed from the
density as $\rhobar=H *\rho$. Here $H({\bfr})$ is a \emph{local}
filter function, which may be of much shorter range than the
potential $G$. Several filter functions are possible, with examples 
$H({\bfr})$ being a $\delta$-function, an exponential  (or
inverse-Helmholtz in 1D) or a top-hat filter. 
\rem{  
The 
normalizing factor of $2 l$ is introduced so that $ \int_{-
\infty}^{+\infty} H({\bfr})
\mbox{d}\,^3x =1$. 
}
Alternatively, for crystal growth applications,  one may assume a filter
function $H({\bfr})$ with zero average, so the mobility does not depend 
on the magnitude of the density, 
only on the relative position of particles. 
\rem{
$ \int_{- \infty}^{+\infty} H({\bfr})\mbox{d}\,^3x =0$. 
} 
 The mathematical analysis
and the reduction property derived in this paper holds,
regardless of the shape of the filter function $H$ and the
potential $G$ is, so long as they are both 
\emph{nice} (\emph{e.g.}, piecewise smooth) functions. 

{\bf Aim of the paper} 
This paper treats the
following continuity equation for the evolution of density,
\begin{equation}\label{rhoeq}
\frac{\partial\rho}{\partial t}
=
-\,
{\rm div}\,{\mathbf{J}}
\quad\hbox{with}\quad
{\mathbf{J}} = 
-\,D\nabla\rhobar -\,\mu(\rhobar)\rho\nabla  \Phi
\,.
\end{equation}
Here $D\ge0$ is a constant, while $\rhobar$ and
$\Phi$ are defined by two convolutions involving,
respectively, the filter function $H$ and the potential $G$,
\begin{equation} 
\label{convols} 
\rhobar=H*\rho
\quad\hbox{and}\quad
\Phi=G*\rho
\,.
\end{equation} 
While model (\ref{rhoeq},\ref{convols}) preserves the sign of $\rho$,
we shall see that it allows the formation of $\delta$-function
singularities even when $D>0$ and in the case of one dimension. The
generality and predictive power of model (\ref{rhoeq},\ref{convols})
can be demonstrated by enumerating a few of its subcases. For example, 
when $H({\bfr})=\delta({\bfr})$ and $G(x)=e^{-|{\bfr}|/l}$ this system
reduces to the generalized chemotaxis equation
\cite{Velazquez2002}. For the choice $G=\delta'({\bfr})$,
$H=\delta'({\bfr})$ one obtains a variant of  the inviscid  Villain model
for MBE evolution \cite{Villain1991}.   If the range of $G$ 
(denoted by $l$) is sufficiently small, one may approximate (at
least formally) the integral operator in $\Phi=G* \rho$ as a
differential operator acting on the density $\rho$, namely,
$\Phi=\rho+l \nabla \cdot ( l \nabla \rho) $.  
As was demonstrated in \cite{MPXB2004}, equation (\ref{rhoeq}) then becomes
a generalization of the viscous Cahn-Hilliard equation, describing
aggregation of domains of different alloys. In the case that $\mu$ is
constant, $H(x-y)=\delta(x-y)\,,$ and $G$ is the Poisson kernel,
then equation (\ref{rhoeq}) becomes the Poisson-Smoluchowski equation for
the interaction of gravitationally attracting particles under Brownian
motion \cite{Ch1939}. 

All four subcases of model (\ref{rhoeq},\ref{convols}) described above
require a singular choice of the smoothing kernels $G$ and $H$. However,
the generalized functions required in these singular choices may be
approximated with arbitrary accuracy by using sequences of nice (for
example, piecewise smooth) functions. We shall concentrate on cases
where the functions $H$ and $G$ remain nice, and derive the
results in this more general, more regular, setting. Thus,
regularization introduces additional
generality, which enables analytical progress and makes
numerical solution of the equations easier. 

{\bf Gradient flow motivation}   Model (\ref{rhoeq},\ref{convols})
may also be motivated from the viewpoint of gradient flows and
thermodynamics. The associated free energy decreases monotonically in time
and remains finite even when the solutions tend to a set of
delta-functions. This energy remains finite for all choices of
$\mu(\rhobar)$. In contrast, the traditional approach in which the
mobility is assumed to depend on the un-smoothed density
$\mu(\rho)$ will fail by allowing the energy to become infinite. This
additional energetic argument reinforces the choice of the regularized 
model in (\ref{rhoeq},\ref{convols}). The calculation will be performed in
arbitrary spatial  dimensions. For the energy to be well-defined, we
assume that the function $G$, describing the interaction between two
particles, is everywhere positive, so the interaction is always
attractive.  In one dimension, we also assume that the kernel function
$G$ is symmetric and in $n$ dimensions that the interaction is central.
These assumptions are physically viable for all the previous subcases. Our
derivation for the free energy will remain valid for an \emph{arbitrary}
filter function $H$. 

\rem{  
\begin{equation}\label{cont-eqn}
\frac{\partial\rho}{\partial t}
=
-\,
{\rm div}\,\mathbf{J}
\end{equation}
\rem{Note that the continuity equation preserves the sign of
the mass density $\rho$.  
According to Darcy's Law, the velocity is obtained 
as the gradient of the pressure,}
The particle flux expression is 
\begin{equation}\label{Darcy}
\mathbf{u}=- \nabla{p}
\,.
\end{equation}
We  shall obtain the relation between the pressure
$p$ and density $\rho$ from classical thermodynamics, as 
\[
p=\frac{\delta E}{\delta \rho}
\,,
\]
for a free energy functional $E[\rho]$. Let us start with the particular case when the mobility is constant $\mu(\rhobar) = \mu_0$. Seeking a situation in which weak
solutions ($\delta$-functions) exist, we specify $E[\rho]$ as a \emph{negative}
 norm of the density, 
\begin{eqnarray}\label{frerg}
E[\rho]=\frac{\mu_0}{2} \int  \rho  \Phi \, d\,^nx
\quad\hbox{with} 
\, \quad \Phi=G*\rho \, \end{eqnarray}
for  some (positive) kernel $G$.  In particular, for 
the kernel $G(x)=\exp (-|x|/ \alpha)$, the energy defines the square of the norm of the density $\|\rho\|^2_{H^{-1}}$ in the Sobolev 
space $H^{-1}$, which remains finite when the solution $\rho$  is either a smooth function  on a set of $\delta$-functions.  
Hence,  in that particular case, the pressure is given by
\begin{eqnarray}\label{p-def}
p=\frac{\delta E}{\delta \rho}=\mu_0\,\rhobar
\,.
\end{eqnarray}
Thus, when the free energy is a negative Sobolev norm, the
thermodynamic approach singles out the case in which the
pressure is linear in a smoothed mass density. For example,
one could choose the free energy $E[\rho]$ to be the
$H^{-1}$ norm of $\rho$. (We note that $\delta$-functions live in
the $H^{-1}$ norm and this motivates us to seek weak solutions
as a sum over $\delta$-functions for this situation, in the
example below.) Using several integration by parts, one can
show that this thermodynamic free energy evolves according to
\begin{equation}\label{ddtnorm}
\frac{d}{dt}E[\rho]
=-\int\rho\,|\mathbf{u}|^2
\,d\,^nx\
\,.
\end{equation}
\
\rem{   
\begin{eqnarray*}\label{ddtnorm}
\frac{d}{dt}E[\rho]
&=&
\frac{\mu_0}{2}\frac{d}{dt}\int \rho\rhobar\,d\,^nx
=
-\mu_0\int\rhobar\,\rm{div}\rho\mathbf{u}
\,d\,^nx
\\&=&
\int(\mu_0\nabla\rhobar)\cdot\rho\mathbf{u}
\,d\,^nx
=\int(\nabla{p})\cdot\rho\mathbf{u}
\,d\,^nx
\\&=&
-\int\rho\,|\nabla{p}|^2
\,d\,^nx
=-
\int\rho\,|\mathbf{u}|^2
\,d\,^nx\
\,.
\end{eqnarray*}
} 
So the rate of decay of thermodynamic free energy defines a
(Riemannian) kinetic energy metric in the transport velocity
\cite{Otto2001}. 

Let us now turn our attention to 
}

We start with equation (\ref{rhoeq}) for mass conservation, in
the case when the mobility $\mu(\rhobar)$ in the particle
flux $\mathbf{J}$ is not constant.  We shall derive a
variant of equation (\ref{rhoeq}) as a gradient flow 
defined by,
\begin{eqnarray}\label{gradflow-def}
\frac{\partial\rho}{\partial t}
=
-\,
{\rm grad}E\big|_\rho
\,,
\hbox{ or }
\Big\langle \frac{\partial\rho}{\partial t}\,,\,\Psi\Big\rangle
=
-\,
\Big\langle \frac{\delta E}{\delta\rho}\,,\,\Psi\Big\rangle
\,,
\end{eqnarray}
where $\langle f\,,\,\Psi\rangle=\int f\Psi\,d^nx$ is the $L^2$
pairing, for a suitable test function $\Psi$ and where $E$ is the
following energy,
\begin{equation} \label{frerg}
E[\rho]
=
D\!\!\int\!\! \rhobar(\log\rhobar-1)\mbox{d}x 
+
\frac{1}{2}\!\!\int\!\!\rho\, \Phi \,\mbox{d}x 
\,,
\end{equation} 
with $\Phi=G*\rho$ and $\rhobar=H*\rho$.
The first term in energy $E[\rho]$ in (\ref{frerg})
decreases monotonically in time for linear diffusion of $\rhobar$. The
second term defines the $H^{-1}$ norm of $\rho$ in 1D for the choice
$G(x)=\exp(-|x|/\alpha)$, and it provides a generalization
of this norm for an arbitrary (but positive and symmetric)
function $G$.  Energy $E[\rho]$ in (\ref{frerg}) remains finite,
even when the  solution for the density $\rho$ concentrates into
a set of $\delta$-functions.

The variation of the free energy
$E[\rho]$ in equation (\ref{frerg}) is
\begin{eqnarray*}\label{Evar1}
\delta E[\rho] 
=
\int(DH*(\log\rhobar) + \Phi) \,\delta\rho\,d\,^nx
\,,
\end{eqnarray*}
where the density variation $\delta\rho$ is chosen in the form,
\begin{eqnarray*}\label{del-rho}
\delta\rho = -\rm{div}\big(\rho\mu(\rhobar) \nabla{
\Psi}\big)
\,,
\end{eqnarray*}
determined by the function $\Psi$, which is assumed to be smooth, 
\cite{footnote}.
After integrating by parts, the corresponding variational derivative is
\rem{
\begin{eqnarray*}\label{Evar2}
\delta E[\rho] 
&=&
\Big\langle \frac{\delta E}{\delta\rho}\,,\, \Psi\Big\rangle
=-\!\!\int\!\!\Psi\,
{\rm div} \Big( 
\rho\mu (\rhobar)\nabla
\big((DH*(\log\rhobar) + \Phi)\big)\Big)
\,d\,^nx 
\,.
\end{eqnarray*}
Consequently, the free energy $E[\rho]$ in equation (\ref{frerg})
produces the following gradient flow:
}
\begin{eqnarray*}
\Big\langle \frac{\delta E}{\delta\rho}\,,\,\Psi\Big\rangle
&=&-
\Big\langle 
\rm{div} \Big( 
\rho\mu (\rhobar)\nabla
\big((DH*(\log\rhobar) + \Phi)\big)\Big)
\,,\,\Psi\Big\rangle
\,.
\end{eqnarray*}
The resulting variant of equation (\ref{rhoeq}) is, cf.
(\ref{gradflow-def})
\begin{equation}\label{rhoeq-var1}
\frac{\partial\rho}{\partial t}
=
-\,
{\rm div}\,{\mathbf{J}}_{m}
\,,\quad 
{\mathbf{J}}_{m} = 
-\,\rho 
\mu (\rhobar)
\big(DH*\frac{1}{\rhobar}\nabla\rhobar
+
\nabla\Phi\big)
\,,
\end{equation}
in which the linear diffusion of density $\rhobar$ is mollified by
smoothing with $H$. Finally obtaining the particle flux $\mathbf{J}$
in equation (\ref{rhoeq}) requires simplifying the first term of
${\mathbf{J}}_{m}$ in (\ref{rhoeq-var1}) to $D\nabla\rhobar$ .  Thus,
only the contribution $-\mu(\rhobar)\rho\nabla\Phi$ to the particle flux 
in equation (\ref{rhoeq}) is variational.



\rem{The pressure $p$ is still defined by the variational
derivative $p=\delta E/\delta \rho$. 
Since this holds for any
test funstion $\Psi$, we have
\begin{eqnarray}\label{HPeqn}
\frac{\partial\rho}{\partial t}
=
\rm{div}(\rho \mu_(\rhobar) \nabla\Phi)
\,,
\quad\hbox{with}\quad
\rhobar=H*\rho 
\,.\end{eqnarray}
} 
{\bf Energetics}
The free energy $E[\rho]$ in (\ref{frerg}) 
decreases monotonically in time under the evolution
equation (\ref{rhoeq-var1}). A direct calculation yields, 
\begin{equation}\label{erg}
\frac{dE[\rho]}{dt}
=
-\int 
\frac{1}{\rho\mu_(\rhobar)}\,|{\mathbf{J}}_m|^2
\,d\,^nx 
\,.
\end{equation}
\rem{
According to this equation, provided $\rho\mu_(\rhobar)>0$,
the rate of decay of free energy $E[\rho]$ given in
(\ref{frerg}) defines a Riemannian  metric in the particle
flux, cf. \cite{Otto2001}.  
}
We shall see that the resulting mass-conserving motion of finite-size
particles governed by (\ref{rhoeq-var1}) leads to a type of
`clumping' of the density $\rho$ into a support set consisting of
$\delta$-functions. One may check that this monotonic decrease of energy
persists for more general functions, including these weak solutions
supported on $\delta$-functions, as discussed below. These weak solutions
are then found to spontaneously emerge in numerical simulations of
equation (\ref{rhoeq}) and dominate  the initial value problem.  

\rem{  
Many of the same results as those described below will also hold for the
nonvariational equation,
\begin{eqnarray}\label{HPeqn}
\frac{\partial\rho}{\partial t}
=
\rm{div}(\rho \nabla P(\rhobar))
\,,
\quad\hbox{with}\quad
\rhobar=G*\rho
\,,\end{eqnarray}
provided $P\,'(\rhobar)>0$.

} 

{\bf 
Dynamics of clumpons}  
We consider motion under equations (\ref{rhoeq},\ref{convols})
in one spatial dimension. Substituting the following singular solution
ansatz 
\begin{eqnarray}\label{Nweak-soln}
\rho(x,t)&=&\sum_{i=1}^Nw_i(t)\delta(x-q_i(t))
\,,
\end{eqnarray}
into the one-dimensional version of either equation (\ref{rhoeq})
or (\ref{rhoeq-var1}) with $D=0$ and integrating the result against a
smooth test function $\phi$ yields 
\rem{\small
\begin{eqnarray*}\label{weak-soln-eqns0}
&&\int\phi\Big[\rho_t-\Big(\rho\,  \mu(\rhobar) (G*\rho)_x\Big)_x
\Big]dx
=
\int\phi(x)\sum_{i=1}^N
\dot{w}_i\,\delta({x-q_i})dx
\\&&
+
\int\phi\,'(x)
\sum_{i=1}^Nw_i\Big(\dot{q}_i+\sum_{j=1}^N w_j(t)\mu \big(\rhobar \big) G\,'(x-q_j)
\Big)\delta({x-q_i})\,dx
\end{eqnarray*}
}
%
a closed set of equations for the
parameters $w_i(t)$ and $q_i(t)$, $i=1,2,\dots,N,$ of the
solution ansatz (\ref{Nweak-soln}). Namely, 
\begin{eqnarray}\label{weak-soln-eqns}
\dot{w}_i(t)=0
\,,\quad
\dot{q}_i(t) = -\sum_{j=1}^N w_j \mu \big(\rhobar \big) G\,'(q_i-q_j)
\,,\end{eqnarray}
where 
$
\rhobar= \sum_{m=1}^N w_m  H(q_j(t)-q_m(t)) 
\label{rhobarsingular} 
$. 
Thus, the density weights $w_i(t)=w_i(0)=w_i$ are preserved for
$i=1,2,\dots,N$, and the corresponding positions $q_i(t)$ follow the
characteristics of the velocity ${\mathbf{u}}=-\mu(\rhobar)\nabla G*\rho$
along the Lagrangian trajectories defined by $x=q_i(t)$. This result holds
in any number of dimensions, modulo changes to allow singular solutions
supported along moving curves in 2D and moving surfaces in 3D.
When $D\ne0$ and div$D\nabla\rhobar\simeq\rho$, the weights $w_i$ decay
essentially exponentially in time, under evolution by
equation (\ref{rhoeq}).

In Fig.~1, we demonstrate that the solutions
(\ref{weak-soln-eqns}) do indeed appear spontaneously in a
numerical simulation of equation (\ref{rhoeq}) with $D=0.02$. Hence,
they are essential in understanding its dynamics. The simulation
started with a smooth (Gaussian) initial condition for density
$\rho$. Almost  immediately, one observes the formation of
several singular \emph{clumpons}, which evolve to collapse
eventually into a single clumpon. Observe that the mass of each
individual clumpon remains almost exactly constant in the
simulations, as required by equation (\ref{weak-soln-eqns}) with $D=0$. Also
note that the masses of two individual clumpons add up when they
collide and ``clump'' together. Consequently, all the mass
eventually becomes concentrated into a single clumpon, whose mass
(amplitude) is the total mass of the initial condition. 
 This simulation shows that for a general system (\ref{rhoeq})
the long-term evolution evolves into the collective motion 
of a \emph{finite} number of individual clumpons. This conclusion
is supported by analysis showing the superposition of clumpons
represented by the singular solution (\ref{Nweak-soln}) is an
invariant manifold of the gradient flow equations (\ref{rhoeq}).


\begin{figure} [h]
\label{fig:collapse} 
\centering 
\includegraphics[width=8cm]{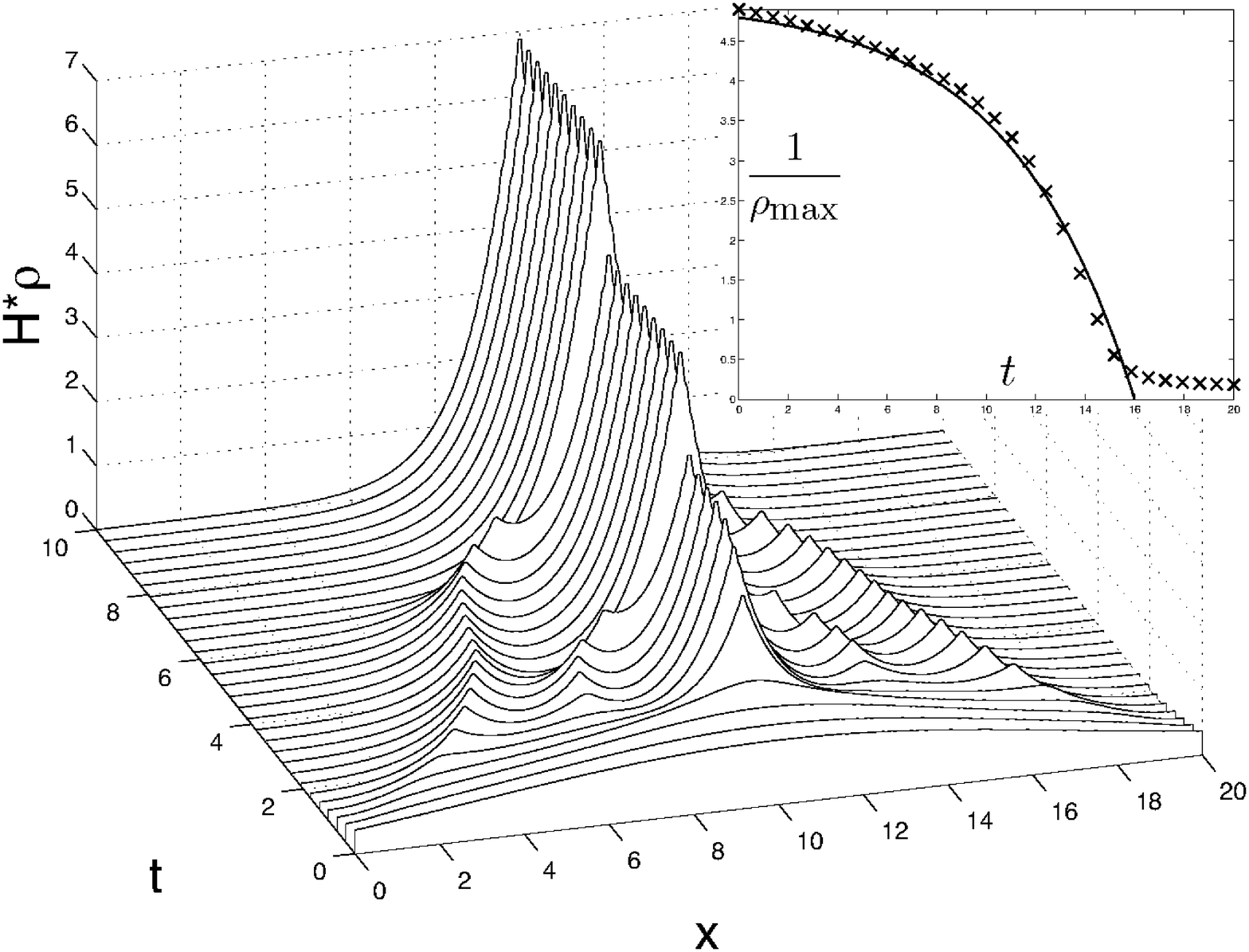} 
 \caption{  Numerical simulation demonstrates the 
emergence of particle clumps, showing formation of density
peaks in a simulation of the initial value problem
for equation (\ref{rhoeq}) using smooth initial conditions for
density with $l=1$, $G(x)=H(x)=\exp (-|x|) $,
$\mu(\rhobar)=1$. The vertical coordinate represents 
$\rhobar=H*\rho$, which remains finite even when the density forms $\delta$-functions. 
 Insert: The
behavior of the inverse amplitude of the maximum peak $1/
\rho_m(t)$ (crosses) versus analytical solution of equation\protect{ (\ref{ricatti})} (solid line). 
 } 
 \end{figure}

{\bf Rate of blow up}
Analysis of the evolution of a density maximum reveals the clumping
process results from the nonlinear instability of the gradient flow in
equation (\ref{rhoeq}) when $D=0$.  For a particular case $G=e^{-|x|/\alpha}$, one may show that for large
enough peak,  a density maximum $\rho_m(t)=\rho(x_m(t),t)$ becomes infinite in finite time.  The motion of the maximum is governed by 
\begin{equation} \label{ricatti}
\frac{d}{d t} \rho_m=\frac{1}{\alpha^2} 
\left( \rho_m^2-\rho_m \Phi(x_m)  \right) \geq \frac{1}{\alpha^2}
\left( \rho_m^2-\rho_m M \right), 
\end{equation} 
where $M=\int \rho \mbox{d} x$ is total mass and we have used the fact that $\Phi$ satisfies
$\Phi-\alpha^2 \Phi_{xx}=\rho$. 
The last inequality holds, because $G \leq 1$ is bounded and
$\rho$ is everywhere positive. Thus, if at any point
the maximum of $\rho$ exceeds the (scaled) value of the total mass, 
then the value of the density maximum $\rho_m(t)$ must diverge in finite
time, which produces $\delta$-functions in finite time. From
(\ref{ricatti}), the density amplitude must diverge as
$\rho_m \simeq \alpha^2/(t_0-t)$. To illustrate this divergence,
we have plotted the comparison between the predicted collapse of
$1/\rho_m$ and numerics  in the insert of
Fig.~1. The formation of 
singularities in Fig.~1 occurs both at the maximum, and elsewhere,
The subsidiary peaks eventually
collapse with the main peak.  

{\bf Competing length scales of $H$ \& $G$}
An interesting limiting case arises, when the scale of non-locality of $H$ is much shorter
than the range of the potential $G$.  Formally, this limit corresponds to 
$H(x)\rightarrow \delta(x)$. In practice, we  may select a
sequence of piecewise smooth functions $H_\epsilon(x)=\exp
(-|x|/\epsilon)/(2 \epsilon)$ which converge weakly to a $\delta$-function.
For each function $H_\epsilon(x)$ in this sequence, no matter
how small (but positive) the value of $\epsilon$, the exact ODE
reduction (\ref{weak-soln-eqns}) still holds. What happens in the
limit of very small epsilon, as
$\epsilon\to0$? 
\rem{  
As is known \cite{Velazquez2002}, for
$\lim_{\epsilon\to0}H_\epsilon(x)=\delta(x)$, equation (\ref{rhoeq})
exhibits weak solutions which are piecewise constants; either $\rho=0$ or
$\rho=1$. 
} 
 In investigating the limit as $\epsilon\to0$, we
performed a sequence of numerical simulations with the function $G(x)=\exp
(-|x|/\alpha)$ being held fixed and $H_\epsilon(x)=\exp
(-x/\epsilon)/(2 \epsilon)$ varying, for a sequence of decreasing
values of the ratio $\epsilon/\alpha$.  This simulation is illustrated
in Fig.~2 for $\epsilon/\alpha=1/10$, which demonstrates
the formation of flat clumps of solutions. The mechanism of this
phenomenon is the following. The vanishing mobility at
$\rhobar=1$ caps the maximal density at $\rhobar=1$ in the
long-term. This leads to the appearance of flat mesas in
$\rhobar(x)$ for large $t$. On the other hand, when $H(x)
\rightarrow \delta(x)$, $\rhobar(x,t) \rightarrow \rho(x,t)$
pointwise in $x$, which forces $\rho(x,t)$ to develop a flat
mesa, or plateau, structure in which the maximum is very close
to unity, as well.  This is precisely what is predicted for the
chemotaxis equation with $H(x)=\delta(x)$ and $\mu(\rho)=1-\rho$
\cite{Velazquez2002}.

\begin{figure} [h]
\label{fig:nocollapse} 
\centering 
\includegraphics[width=8cm]{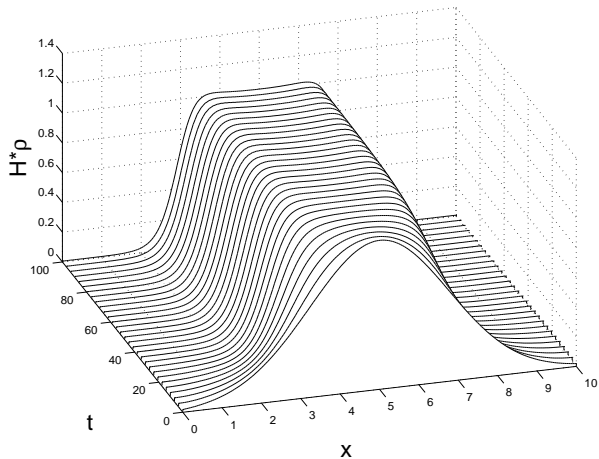} 
\caption{ 
Evolution of a Gaussian initial
condition for $\rho$(x,0) with $\mu(\rhobar)=1-\rhobar$ and
$H=\exp(-|x|/\epsilon)/(2 \epsilon)$ where $\epsilon=\alpha/10$. The
solution quickly forms a plateau of maximal possible density
($\rho_{\hbox{\small max}}=1$).  }
\end{figure}

In the limit $H \rightarrow \delta$,  model (\ref{rhoeq},\ref{convols}) 
recovers ordinary diffusion of local density. 
This is a singular limit,  because it increases the order of
the differentiation in the equation. Since ordinary
diffusion  is known to prevent collapse in one dimension \cite{Ho2003}, this singular
limit should be of considerable interest for further analysis.  

{\bf Conclusion and Open Problems} 
A new model was proposed and analyzed for the
collective aggregation of finite-size particles driven by the force of mutual attraction.
Starting from smooth initial conditons, the solution
for the particle density in this model was found to collapse into a set of
delta-functions (clumps), and the evolution equations for the dynamics of these clumps was
computed analytically. The energy derived for this model is well defined 
even when density is  supported on
$\delta$-functions. The mechanism for the formation of these
$\delta$-function clumps is the nonlinear instability governed by the
Ricatti equation (\ref{ricatti}), which causes the magnitude of any density
maximum to grow without bound in finite time.
\rem{ 
In this letter, we proposed and analyzed a new model for the
collective aggregation of finite-size particles driven by the force 
of mutual attraction. The model exhibits the formation 
of singularities. 
} 
At first sight, it may seem that the emergence of
$\delta$-function peaks in the solution might be undesirable and
perhaps should be avoided. However, these $\delta$-functions
may be understood as clumps of matter, and the model guarantees that
any solution eventually ends up as a set of these clumps.
Consequently, further collective motion of these clumps may be
predicted using a (rather small-dimensional) system of ODEs,
rather than dealing with the full non-local PDEs.  
The question of how many clumps arise from a given initial condition 
remains to be considered. One may conjecture that clump formation is
extensive; so that each clump forms from the material within the range of
the potential $\Phi$, determined by $G$. On a longer time scale, the clumps
themselves continue to aggregate, as determined by the collective
dynamics (\ref{weak-soln-eqns}) of weak solutions (\ref{Nweak-soln}). This
clump dynamics is also a gradient flow; so that eventually only one clump
remains. 

An analogy to point vortex solutions of Euler's equations for
two-dimensional ideal hydrodynamics may be drawn here. Prediction
of motion of inviscid fluid is governed by a
set of nonlinear PDE in which pressure introduces non-locality.
A drastic  simplification of motion occurs, when all the 
vorticity is concentrated in delta-functions (point vortices) 
\cite{Saffman-book}. The motion of point
vortices lies on a singular invariant manifold: if
started with a set of point vortices, the fluid structure will
remain a set of point vortices. However, a smooth initial
condition for vorticity \emph{does not} split into point
vortices under the Euler motion. In the present model, though,
the physical attraction drives any initial conditions towards a
set of delta-functions, so one is \emph{guaranteed} to obtain
effectively finite-dimensional behavior in the system after a
rather short initial time.  

Because two scales are present in the smoothing functions $H$ and $G$, the
effects of boundary conditions warrant further study. For example, 
 the $H \rightarrow \delta$ limit should also allow
formation of boundary layers. Here, these boundary issues were avoided
by using periodic boundary conditions.  However, it is natural in physical
situations to apply, e.g., Neumann boundary conditions to the particle flux
$\mathbf{J}$.  Thus, the issue of boundary conditions may
become important in a more physically realistic setting. 
\rem{
Finally, we remark that singular solutions (\ref{Nweak-soln}) exist only
for $\tau=0$ and $D=0$ in the Keller-Segel equations (\ref{Keller-Segel}).
Study of the effect of $\tau\neq 0$ on these solutions remains another
interesting open direction for future research, beyond the scope of the
present work.
}

{\bf Acknowledgments.}  We thank P. Constantin, B. J. Geurts, J.~Krug and E. S. Titi for
encouraging discussions and correspondence about this work. We are
grateful for partial support by US DOE, under contract W-7405-ENG-36 for
Los Alamos National Laboratory, and Office of Science ASCAR/AMS/MICS.


\begin{thebibliography}{99}
\bibitem{footnote} 
These gradient flow properties of the smooth solutions are also preserved
for the weak solutions.
\end{thebibliography}
\end{document}